\documentclass[twocolumn,10pt,aip,jcp,amsmath,amssymb]{revtex4-1}
\usepackage[utf8]{inputenc}
\usepackage{amsmath}
\usepackage{graphicx}   
\usepackage{dcolumn}    
\usepackage{bm}         
\usepackage{hyperref}
\usepackage{color}

\newcommand{\wt}{\widetilde}
\newcommand{\ninf} {$N\to\infty$ }
\newcommand{\ninc} {$N\to\infty$, }

\newcommand{\jcp}{J.\ Chem.\ Phys.\ }

\newcommand{\jpca}{J.\ Phys.\ Chem.\ A }
\newcommand{\jpcc}{J.\ Phys.\ Chem.\ A }

\newcommand{\jpcl}{J.\ Phys.\ Chem.\ Lett.\ }

\newcommand{\eqn}[1]{Eq.~(\ref{#1})}
\newcommand{\Eqn}[1]{Equation~(\ref{#1})}
\newcommand{\eqnn}[2]{Eqs.~(\ref{#1}) and (\ref{#2})}

\newcommand{\eqnt}[2]{Eqs.~(\ref{#1})-(\ref{#2})}

\newcommand{\eb}{e^{-\beta \hat H}}

\newcommand{\ebN}{e^{-\beta_N \hat H}}

\newcommand{\etf}{e^{-i \hat H t/\hbar}}
\newcommand{\etb}{e^{i \hat H t/\hbar}}

\newcommand{\bra}[1]{\langle #1 |}
\newcommand{\ket}[1]{| #1 \rangle}
\newcommand{\kb}[1]{\ket{#1}\bra{#1}}

\newcommand{\ddp}[2]{\frac{\partial #1}{\partial #2}}

\newcommand{\tr}{ {\rm Tr} }

\newcommand{\no}{\nonumber}

\newcommand{\ola}{\overleftarrow}
\newcommand{\ora}{\overrightarrow}

\newcommand{\bp}{ {\bf p} }

\newcommand{\bq}{ {\bf q} }

\newcommand{\bz}{ {\bf z} }
\newcommand{\bDelta}{ {\bf \Delta} }

\newcommand{\bQ}{ {\bf Q} }

\newcommand{\tph}{\frac{1}{2\pi\hbar}}

\newcommand{\tphN}{\frac{1}{(2\pi\hbar)^N}}

\newcommand{\inti}{\int_{-\infty}^{\infty}}

\newcommand{\eql}[1]{\label{eq:#1}}

\begin{document}

\title{Boltzmann-conserving classical dynamics in quantum time-correlation functions: `Matsubara dynamics'} 
\author{Timothy J.~H.~Hele} 
\noaffiliation
\author{Michael J.~Willatt}
\noaffiliation
\author{Andrea~Muolo}
\email[Current address: Lab. für Physikalische Chemie, ETH Zürich, CH-8093 Zürich, Switzerland]{}
\noaffiliation
\author{Stuart C.~Althorpe}
\email[Corresponding author: ]{sca10@cam.ac.uk}
\noaffiliation
\affiliation{\mbox{Department of Chemistry, University of Cambridge, Lensfield
Road, Cambridge, CB2 1EW, UK.}}
\date{\today}

\begin{abstract}
We show that a single change in the derivation of the linearized
semiclassical-initial value representation (LSC-IVR or `classical Wigner
approximation')  results in a classical dynamics which conserves the quantum
Boltzmann distribution. We rederive the (standard) LSC-IVR approach by writing
the (exact) quantum time-correlation function in terms of the normal modes of a
free ring-polymer (i.e.\ a discrete imaginary-time Feynman path), taking the
limit that the number of polymer beads $N\rightarrow\infty$, such that the
lowest normal-mode frequencies take their `Matsubara' values. The change we
propose is to truncate the quantum Liouvillian, not explicitly in powers of
$\hbar^2$ at $\hbar^0$ (which gives back the standard LSC-IVR approximation),
but in the normal-mode derivatives corresponding to the lowest Matsubara
frequencies. The resulting `Matsubara' dynamics is {\em inherently classical}
(since all terms ${\cal O}(\hbar^2)$ disappear from the Matsubara Liouvillian in
the limit $N\rightarrow\infty$), and conserves the quantum Boltzmann
distribution because the Matsubara Hamiltonian is symmetric with respect to
imaginary-time translation.  Numerical tests show that the Matsubara
approximation to the quantum time-correlation function converges with respect to
the number of modes, and gives better agreement  than LSC-IVR with the exact
quantum result. Matsubara dynamics is too computationally expensive to be
applied to complex systems, but its further approximation may lead to practical
methods. \textit{Copyright (2015) American Institute of Physics. This article
    may be downloaded for personal use only. Any other use requires prior
    permission of the author and the American Institute of Physics.  The
    following article appeared in the Journal of Chemical Physics, 142, 134103
(2015) and may be found at http://dx.doi.org/10.1063/1.4916311}
\end{abstract}

\maketitle 

\section{Introduction}

Dynamical properties at thermal equilibrium are of central importance to chemical physics.\cite{green,daan} Sometimes these properties can be simulated adequately by entirely classical means. But there are plenty of cases, e.g.\ the spectrum of liquid water,\cite{rpmd_water,lsc-ivr_water,marx_water} hydrogen-diffusion on metals,\cite{aedes,yury} and proton/hydride-transfer reactions,\cite{hammes1,tmiller1,mano_borg,makri1,shi,rommel} for which one needs to evaluate time-correlation functions of the form 
\begin{align}
 {1\over Z}\overline C_{AB}(t) = {1\over Z}\tr\left[\eb \hat A \etb \hat B \etf\right] \label{poodle}
\end{align}
(where $Z$ is the partition function,\cite{def}  $\beta \equiv 1/k_{\rm B}T$, Tr indicates a complete sum over states, and the other notation is defined in Sec.~II). Such time-correlation functions are already approximate, since they employ
the quantum Boltzmann distribution $\eb/Z$ in place of 
 the exact quantum-exchange statistics; but  this approximation is usually adequate (since the thermal wavelength is typically much smaller than the separations between identical particles). What is less well understood is the extent to which such functions can be further approximated by replacing the exact quantum dynamics by classical dynamics (whilst retaining the quantum Boltzmann statistics). 

The standard way to make this approximation is to use the linearized semiclassical-initial value representation (LSC-IVR, sometimes called the `classical Wigner' approximation),\cite{lsc-ivr_water,miller_rev,wang_miller,miller_molphys,liumill,poulsen,coker1,coker2,shi_geva,hillery,heller,Shig2,Liurev,borgis} in which the quantum Liouvillian is expanded as a power series in $\hbar^2$, then truncated at~$\hbar^0$. Miller\cite{miller_rev,miller_molphys} and later Shi and Geva\cite{shi_geva} showed that this approximation is equivalent to linearizing the displacement between forward and backward Feynman paths in the exact quantum time-propagation, which removes the coherences, thus making the dynamics classical. The LSC-IVR retains the Boltzmann quantum statistics inside a Wigner transform,\cite{hillery} is exact in the zero-time, harmonic and  high-temperature limits, and has been developed into a practical method by several authors.\cite{liumill,Shig2,poulsen,Liurev,borgis} However, it has a serious drawback: the classical dynamics does not in general preserve the quantum Boltzmann distribution, and thus the quality of the statistics deteriorates over time.

A number of methods have been developed to get round this problem, all of which appear to some extent to be ad hoc.  Some of these methods are obtained by replacing the plain Newtonian dynamics in the LSC-IVR by an effective (classical) dynamics which preserves the Boltzmann distribution.\cite{liu1,liu2,poul1} Others, such as the popular centroid molecular dynamics (CMD)\cite{jang,rossky} and ring-polymer molecular dynamics (RPMD),\cite{rpmd_water,yury,tmiller1,mano_borg,craig1,craig2,craig3,markman,rpmd_rev,tmiller3,tmiller4,kinetic,nandini1,yury2,huo1,huoaoiz,rossky2,trpmd,
jeremy,tim1,tim2,tim3,tim_thesis,jeremy_new,conan2}  are more heuristic (and still not fully understood) but have the advantage that they can be implemented directly in classical molecular dynamics codes. An intriguing property of CMD and RPMD is that, for some model systems (e.g.~the  one-dimensional quartic oscillator\cite{craig1,rossky}),  these methods give better agreement than LSC-IVR with the exact quantum result, even though, like LSC-IVR, they completely neglect real-time quantum coherence. 

This last point suggests that the failure of LSC-IVR to preserve the quantum Boltzmann distribution may arise, not from its neglect of quantum coherence, but from its inclusion of `rogue' components in the classical dynamics.  The present paper develops a theory that supports this speculation. We isolate a core, Boltzmann-conserving, classical dynamics, which we  call `Matsubara dynamics' (for reasons to be made clear). Matsubara dynamics is far too expensive to be used as a practical method, but is likely to prove useful in understanding methods such as CMD and RPMD, and perhaps in developing new approximate methods.

The paper is structured as follows. Section II gives key background material including the well known `Moyal series' derivation of the LSC-IVR. Section III re-expresses the standard results of Sec.~II in terms of `ring-polymer' coordinates, involving points along the imaginary-time path-integrals that describe the quantum Boltzmann statistics. Section IV gives the new results, showing that smooth Fourier-transformed combinations of the ring-polymer coordinates lead to an {\em inherently classical} dynamics which is {\em quantum-Boltzmann-conserving}. Section V reports numerical tests on one-dimensional models. Section VI concludes the article.

\section{Background theory}

We start by defining the terms and notation to be used in classical and quantum Boltzmann time-correlation functions (IIA and IIB), and by writing out the standard Moyal-series derivation of the LSC-IVR (IIC).
\label{sec:rev}
\subsection{Classical correlation functions}
\label{ssec:clascor}
Without loss of generality, we can consider an $F$-dimensional Cartesian system with position coordinates $\bq\equiv q_1, \ldots, q_F$, momenta $\bp$, mass $m$ and Hamiltonian
\begin{align}
 H(\bp,\bq) = \frac{\bp^2}{2m} + V(\bq)\label{classf}
\end{align}
The thermal time-correlation function between observables $A(\bp,\bq)$, $B(\bp,\bq)$ is then
\begin{align}
 c_{AB}(t) = & \tphN \int d\bp \int d\bq \ e^{-\beta H(\bp,\bq)} \no\\
             & \times A(\bp,\bq) B(\bp_t,\bq_t) \label{dogs}
\end{align}
where  $\int d\bp \equiv \inti dp_1\ldots\inti dp_F$ (and similarly for ${\bf q}$), and
$\bp_t \equiv \bp_t(\bp,\bq,t)$ and $\bq_t \equiv \bq_t(\bp,\bq,t)$ are the momenta and positions after the classical dynamics has evolved for a time~$t$.

Alternatively, we can express $B(\bp_t,\bq_t)$ as a function of the initial phase-space coordinates~$({\bf p},{\bf q})$:
\begin{align}
B(\bp_t,\bq_t)\equiv B[\bp_t(\bp,\bq,t),\bq_t(\bp,\bq,t)]\equiv B({\bf p},{\bf q},t)
\end{align}
such that 
\begin{align}
 c_{AB}(t) = & \tphN  \int d\bp \int d\bq \ e^{-\beta H(\bp,\bq)} \no\\
             & \times A(\bp,\bq) B({\bf p},{\bf q},t) \no\\
           = & \tphN  \int d\bp \int d\bq \ e^{-\beta H(\bp,\bq)} \no\\
             & \times A(\bp,\bq) e^{{\cal L}_Ft}B({\bf p},{\bf q},0)\label{funf}
\end{align}
where the (classical) Liouvillian ${\cal L}_F$ is\cite{zwanzig}
\begin{align}
{\cal L}_F = {1\over m}{\bf p}\cdot{\bf \nabla_q}-V({\bf q})\ola \nabla_\bq \cdot \ora \nabla_\bp \label{arrows}
\end{align}
with
\begin{align}
 \nabla_\bq = 
 \left(\!
 \begin{array}{c}
  \ddp{}{q_1} \\
  \vdots \\
  \ddp{}{q_F}
 \end{array}
 \!\right)
\end{align}
and the arrows indicate the direction in which the derivative operator is applied (and the backward arrow indicates that the derivative is taken only of $V({\bf q})$---not of any terms that may precede  $V({\bf q})$ in any integral). \Eqn{funf} is less practical than \eqn{dogs} (which propagates individual trajectories rather than the distribution function $B({\bf p},{\bf q},t)$) but is better for comparison with the exact quantum expression.

An essential property of the dynamics is that it preserves the (classical) Boltzmann distribution, which follows because $H({\bf p},{\bf q})$ is a constant of the motion. As a result, we can rearrange \eqn{funf} as
\begin{align}
 c_{AB}(t) = & \tphN \int d\bp \int d\bq \ e^{-\beta H(\bp,\bq)} \no\\
             & \times \left[e^{-{\cal L}_Ft}A(\bp,\bq)\right] B({\bf p},{\bf q},0)
\end{align}
showing that $c_{AB}(t)$  satisfies
\begin{align}\label{dbal}
 c_{AB}(t) =c_{BA}(-t)
 \end{align}
which is the {\em detailed balance} condition.

\subsection{Quantum correlation functions}
For clarity of presentation, we will derive the results in Secs.~III and IV for a one-dimensional quantum system with Hamiltonian  $\hat H = \hat T + \hat V$,  kinetic energy operator $\hat T = \hat p^2/2m$,  potential energy operator $\hat V = V(\hat q)$, position and momentum operators ${\hat q,\hat p}$, and mass $m$. However, the results we derive in Secs.~III and IV are applicable immediately to systems with any number of dimensions (see Sec.~IV.D).

 The simplest form of quantum-Boltzmann time-correlation function is that given in \eqn{poodle}, but  ${\overline C}_{AB}(t)$ is difficult to relate to the classical time-correlation function $c_{AB}(t)$,  because it does not satisfy Eq.~(\ref{dbal}) and is not in general real. We therefore use the Kubo-transformed  time-correlation function\cite{craig1}
\begin{align}
 C_{AB}(t) = \tr\left[K_\beta({\hat A})\, \etb \hat B \etf \right] \label{kubo}
\end{align}
with
\begin{align}
K_\beta({\hat A})=\frac{1}{\beta}\int_0^{\beta}d\lambda \ e^{-\lambda \hat H} \hat A e^{-(\beta - \lambda) \hat H}\label{commy}
\end{align}
This function gives an equivalent description of the dynamics to ${\overline C}_{AB}(t)$, to which it is related by a simple Fourier-transform formula.\cite{craig1}

It is easy to show (by noting that $e^{-\lambda \hat H}$ and $\etf$ commute in \eqn{kubo}) that ${ C}_{AB}(t)$ satisfies the detailed balance relation
\begin{align}
 C_{AB}(t) =   C_{BA}(-t) 
\end{align}
This relation also ensures that ${ C}_{AB}(t)$ is real (since reversing the order of operators in the trace gives $C_{AB}(t) =   C_{BA}^*(-t)$).

The $t=0$ limit of $C_{AB}(t)$ can be expressed\cite{craig1} in terms of a classical Boltzmann distribution over an extended phase space of `ring-polymers'.\cite{chandler_wolynes,parrinello,ceperley,charu} When 
${\hat A}$ and ${\hat B}$ are functions ${A}(\hat q)$ and ${B}(\hat q)$ of the position operator $\hat q$, the ring-polymer expression is
\begin{align}
C_{ AB}(0) = & \lim_{N\to\infty}{1\over (2\pi\hbar)^N} \int\!d{\bf p}\int\!d{\bf q}\ \no\\
             & \times A({\bf q})B({\bf q}) e^{-\beta_N R_N({\bf p},{\bf q})}\label{polly}
\end{align}
where $\beta_N=\beta/N$, $\int\!d{\bf p}\equiv\int_{-\infty}^\infty d{p_1}\dots \int_{-\infty}^\infty d{p_N}$ and similarly for
$\int\!d{\bf q}$, and
\begin{align}
 A(\bq) =& \frac{1}{N}\sum_{i=1}^N A({q}_i),\qquad B(\bq) = \frac{1}{N} \sum_{i=1}^N B({q}_i)\label{aaa}\\
 R_N({\bf p},{\bf q})=&T_N({\bf p},{\bf q})+U_N({\bf q})\\
T_N({\bf p},{\bf q}) =& {{\bf p}^2\over 2 m}+{m\over2(\beta_N\hbar)^2} \sum_{i=1}^N(q_{i+1}-q_i)^2\label{tn}\\
U_N({\bf q})=&\sum_{i=1}^NV(q_i)\label{UN}
\end{align}
Similar expressions can be obtained when ${\hat A}$ and ${\hat B}$ depend on 
the momentum operator (by inserting position-momentum Fourier-transforms).
To avoid confusion, we emphasise that \eqn{polly} is exact at $t=0$, and that we do {\em not} assume that the ring-polymer Hamiltonian $R_N({\bf p},{\bf q})$ generates the dynamics at $t>0$.

\subsection{The LSC-IVR approximation}
\label{ssec:lincor}
\subsubsection{The Wigner-Moyal series}
To derive the LSC-IVR approximation to $C_{AB}(t)$, we follow ref.~\onlinecite{hillery}, expanding the exact quantum Liouvillian in powers of $\hbar^2$.
We start by rewriting \eqn{kubo} as
\begin{align}
  C_{AB}(t)  = & \int_{-\infty}^\infty dq \int_{-\infty}^\infty d\Delta \no\\ 
               & \times \bra{q-\Delta/2} K_\beta({\hat A})\ket{q+\Delta/2}\no\\
               & \times\bra{q+\Delta/2} \etb \hat B \etf \ket{q-\Delta/2} 
\end{align}
then insert the momentum identity
\begin{align}
\delta(\Delta-\Delta')={1\over 2\pi\hbar}\int_{-\infty}^\infty dp \ e^{ip(\Delta-\Delta')/\hbar}
\end{align}
to obtain
\begin{align}
C_{AB}(t) = & \tph \int_{-\infty}^\infty dq \int_{-\infty}^\infty dp \no\\
            & \times [K_\beta({\hat A})]_{\rm W}(p,q)\,  [\hat B(t)]_{\rm W}(p,q) \label{cabw}
\end{align}
where the Wigner transforms of $\hat A$ and $\hat B$ are given by
\begin{align}
 [K_\beta({\hat A})]_{\rm W}(p,q) = & \int_{-\infty}^\infty d\Delta e^{ip\Delta/\hbar} \no\\
                                    & \times \bra{q-\Delta/2} K_\beta({\hat A}) \ket{q+\Delta/2}
\end{align}
and
\begin{align}
 [\hat B(t)]_{\rm W}(p,q) = & \int_{-\infty}^\infty d\Delta \ e^{ip\Delta/\hbar} \no\\
                            & \times \bra{q-\Delta/2} \etb \hat B \etf \ket{q+\Delta/2}.
\end{align}
(and note that we will often  suppress the $(p,q)$ dependence of $[K_\beta({\hat A})]_{\rm W}$ and $[\hat B(t)]_{\rm W}$).

We then differentiate \eqn{cabw} with respect to $t$, 
\begin{align}
{d C_{AB}(t)\over dt} = & \tph  \int_{-\infty}^\infty dq \int _{-\infty}^\infty dp \no\\
                        & \times[K_\beta({\hat A})]_{\rm W}  \left[\frac{i}{\hbar}[\hat H, \hat B(t)]\right]_{\rm W}\label{difft}
\end{align}
and expand  the potential-energy operator in the commutator in powers of $\Delta$ to obtain
\begin{align}
 \left[\frac{i}{\hbar}[\hat H, \hat B(t)]\right]_{\rm W} = & \int_{-\infty}^\infty d\Delta \ e^{ip\Delta/\hbar} \no\\
                                                           & \times {\hat \ell}\bra{q-\Delta/2} \hat B(t) \ket{q+\Delta/2}\label{begetter}
 \end{align}
 with
 \begin{align}
 {\hat \ell}=\frac{i\hbar}{m} \ddp{}{q}\ddp{}{\Delta}-{2i\over\hbar}\sum_{\lambda=1,{\rm odd}}^\infty{1\over \lambda!}{\partial^\lambda V(q) \over\partial q^\lambda}
   \left({\Delta\over 2}\right)^\lambda \label{lambs}
  \end{align}
  Noting that each power of $\Delta$ can be generated by an application of ${-i\hbar\partial/\partial p}$, we then obtain   
\begin{align}
  \frac{dC_{AB}(t)}{dt} = & \tph  \int_{-\infty}^\infty\! dq \int_{-\infty}^\infty\!  dp \no\\
                       & \times [K_\beta({\hat A})]_{\rm W} \,{\hat L}[\hat B(t)]_{\rm W} \eql{cabmoy}
\end{align}
with
 \begin{align}
 \hat L =& \frac{p}{m} \frac{\partial}{\partial q}-\sum_{\lambda=1,{\rm odd}}^\infty {1\over \lambda!}\left( {i\hbar\over 2}\right)^{\lambda-1}{\partial^\lambda V(q) \over\partial q^\lambda}{\partial^\lambda \over\partial p^\lambda}\label{lexp}
\end{align}
This is the Moyal expansion of the quantum Liouvillian in powers of $\hbar^2$. If all terms are included in the series, then the application of $\hat L$  generates the exact quantum dynamics (as is easily proved by working backwards through the derivation just given). A compact representation of $\hat L$, which will be useful later on is, 
\begin{align}
  \hat L =& \frac{p}{m} \frac{\partial}{\partial q} - V(q) \frac{2}{\hbar} \sin\!\left(\frac{\ola \partial}{\partial q}\frac{\hbar}{2}\frac{\ora \partial}{\partial p} \right).
\end{align}
where the arrows are defined in the same way as in \eqn{arrows}

\subsubsection{Approximating the dynamics}

To obtain the LSC-IVR one notes that \eqn{lexp} can be written
\begin{align}
 {\hat L} =
  & \mathcal{L} + \mathcal{O}(\hbar^2)
\end{align}
where $\mathcal{L}$ is the classical Liouvillian
\begin{align}
\mathcal{L} =\frac{p}{m} \frac{\partial}{\partial q}-  \frac{\partial V}{\partial q} \frac{\partial}{\partial p} 
\end{align}
and then truncates ${\hat L}$ at $\hbar^0$. The  LSC-IVR thus amounts to replacing the quantum dynamics by classical dynamics, such that $C_{AB}(t)$ is approximated by
\begin{align}
  C_{AB}^{\rm W}(t) = & \int_{-\infty}^\infty\! dq \int_{-\infty}^\infty\! dp \no\\ 
                      & \times [K_\beta({\hat A})]_{\rm W}(p,q) \,e^{\mathcal{L}t}[\hat B(0)]_{\rm W}({p},{q})
\end{align}
or equivalently
\begin{align}
  C_{AB}^{\rm W}(t) = & \int_{-\infty}^\infty\! dq \int_{-\infty}^\infty\! dp \no\\
                      & \times [K_\beta({\hat A})]_{\rm W}(p,q) \,[\hat B(0)]_{\rm W}({p}_t,{q}_t)
\end{align}
where $(p_t,q_t)$ are the (classical) position and momentum at time $t$ of a trajectory initiated at $(p,q)$ at $t=0$.

Physical insight into the LSC-IVR is obtained by going back to \eqn{lambs}, and noting that truncating $\hat L$ at $\hbar^0$ is equivalent to truncating ${\hat l}$ at $\Delta$. Since $\Delta$ is the difference between the origin of a forward path that terminates at $z$ (at time $t$) and the terminus of a backward path that originates at $z$, it follows that truncating at $\Delta$ is equivalent to linearizing the difference between the forward and backward Feynman paths at each time-step. Hence the neglect of terms ${\cal O}(\hbar^2)$ is valid if the forward and backward paths are very close together, in which case there are no coherence effects, and the dynamics becomes classical. The LSC-IVR is thus exact at $t=0$  (where the paths become infinitessimally short), in the harmonic limit (where the are no terms ${\cal O}(\hbar^2)$ in $\hat L$), and in the high temperature limit (where fluctuations in $\Delta$ efficiently dephase).\cite{summar}

Despite these positive features, LSC-IVR suffers from the major drawback of not preserving the quantum Boltzmann distribution (except in one of the special limits just mentioned), since in general
\begin{align}
\mathcal{L} [\eb]_{\rm W} \neq 0 \eql{moyzero}
\end{align}
As a result, 
\begin{align}
 C_{AB}^{\rm W}(t) & \ne C_{BA}^{\rm W}(-t)
 \end{align}
i.e.\ the LSC-IVR does not satisfy detailed balance. In Secs.~III-V we will
investigate why this is so.


\section{Ring-polymer coordinates}

We now recast the standard expressions of Sec.~II in terms of ring-polymer coordinates. No new approximations are obtained, but the ring-polymer versions of these expressions are needed for use in Sec.~IV, where they will be used to derive the quantum-Boltzmann-conserving `Matsubara' dynamics.

\subsection{Ring-polymer representation of Kubo-transformed time-correlation functions}

\subsubsection{Exact quantum time-correlation function}
Following ref.~\onlinecite{tim1} (see also refs.~\onlinecite{shi_geva} and \onlinecite{nandini1}), we define the ring-polymer quantum time-correlation function to be
\begin{align}
 C_{ AB}^{[N]}(t) = & \int d\bq \int d\bDelta \int d\bz \ A(\bq) B(\bz) \no\\
 & \times \prod_{l=1}^N \bra{q_{l-1} - \Delta_{l-1}/2} \ebN \ket{q_{l}+\Delta_l/2} \no\\
 & \qquad \times \bra{q_{l}+\Delta_l/2} \etf \ket{z_l} \no\\
 & \qquad \times \bra{z_l} \etb \ket{q_{l} - \Delta_{l}/2} \label{monkey}
\end{align}
where the functions $A({\bf q})$ and $B({\bf z})$ (with ${\bf z}$ in place of ${\bf q}$) are defined in \eqn{aaa} 
(and we have assumed that $\hat A$ and $\hat B$ are functions of 
 position operators to simplify the algebra---see Sec.~IVD).
 It is easy to show (by noting that $N\!-\!1$ of all the forward-backward propagators are identities, and that the sums in $A({\bf q})$ and $B({\bf z})$ become integrals in the limit $N\to\infty$) that
\begin{align}
C_{ AB}(t) = \lim_{N\rightarrow \infty} C_{ AB}^{[N]}(t)\label{brolly}
\end{align} 
In other words, \eqn{monkey} in the limit $N\to\infty$ is just an alternative way of writing out
 the standard Kubo-transformed time-correlation function $C_{ AB}(t)$. The advantage of \eqn{monkey} is that it emphasises the symmetry of the entire path-integral expression with respect to cyclic permutations of the coordinates $q_l\to q_{l+1}$ (see Fig.~1); this symmetry is otherwise hidden in the conventional expression for $C_{ AB}(t)$
[\eqn{kubo}]. 

\begin{figure}[t]
    \begin{center}
        \centering
        \includegraphics[scale=0.65]{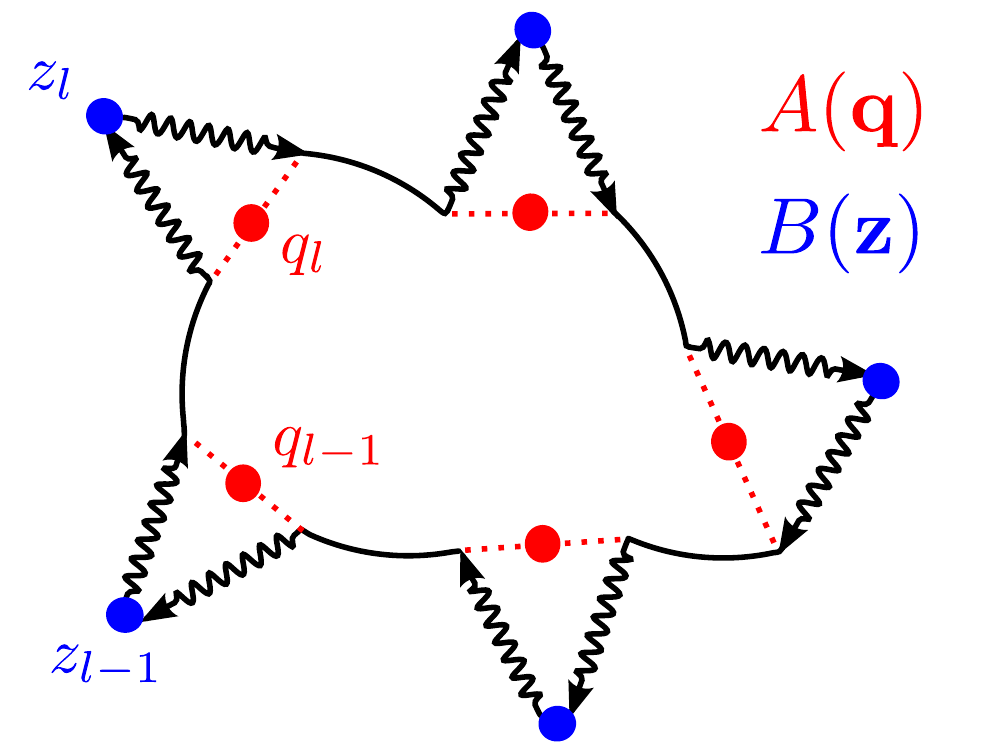}
        \caption{Schematic diagram showing the structure of the (exact)
            Kubo-transformed quantum time-correlation function when represented
            in ring-polymer coordinates as in \eqn{monkey}. The red and blue
            dots represent the coordinates $q_l$ and $z_l$; solid lines
            represent stretches of imaginary time of length $\beta_N\hbar$;
            arrows represent forward-backward propagations in real time.}
        \label{figure1}
    \end{center}
\end{figure}

\subsubsection{Ring-polymer representation of the LSC-IVR}

It is straightforward to derive the LSC-IVR approximation  from \eqn{monkey} by generalizing the steps in Sec.~IIC. We insert an identity
\begin{align}
\delta(\Delta_l-\Delta_l')={1\over 2\pi\hbar}\int_{-\infty}^\infty dp_l \ e^{ip_l(\Delta_l-\Delta_l')/\hbar}
\end{align}
for each value $l=1,\dots,N$, to obtain

\begin{widetext}
\begin{align}
 C_{ AB}^{[N]}(t) = & {1\over(2\pi\hbar)^N}\int d\bq \int d\bp \left[ e^{-\beta{\hat H}} {\hat A}\right]_{\overline  N}({\bf p},{\bf q}) \,\left[ {\hat B}(t)\right]_{N}({\bf p},{\bf q})\label{disgrace}
\end{align}
where 
\begin{align}
\left[ e^{-\beta{\hat H}} {\hat A}\right]_{\overline N} ({\bf p},{\bf q})= & \int d\bDelta  \ A(\bq) 
  \prod_{l=1}^N \bra{q_{l-1} - \Delta_{l-1}/2} \ebN \ket{q_{l}+\Delta_l/2} e^{ip_l\Delta_l/\hbar}
\label{wig1}
\end{align}
and
\begin{align}
\left[ {\hat B}(t)\right]_{N}({\bf p},{\bf q}) = & \int d\bDelta  \int d{\bf z}  \ B(\bz)
 \prod_{l=1}^N \bra{q_{l}-\Delta_l/2} \etf \kb{z_l} \etb \ket{q_{l}+ \Delta_{l}/2} e^{ip_l\Delta_l/\hbar}
 \label{wig2}
\end{align}
\end{widetext}
are generalized Wigner transforms (and we will often suppress the dependence on $({\bf p},{\bf q})$ in what follows).
Note that $[\cdot ]_N$ and $[\cdot ]_{\overline N}$ have different forms: $[\cdot ]_N$ is a sum of products of one-dimensional Wigner transforms, whereas $[\cdot ]_{\overline N}$ is more complicated, with each product coupling variables in $l$ and $l+1$.\cite{alter} 
 Note that since we have specified that ${\hat B}$ is a function of just the position operator (in order to simplify the algebra---see Sec.~IVD), it follows that
\begin{align}
\left[ {\hat B}(0)\right]_{N}({\bf p},{\bf q}) = B({\bf q})
 \eql{mgk}
\end{align}

The next step is to obtain the ring-polymer representation of the (exact) quantum Liouvillian, which involves a straightforward generalization of \eqnt{difft}{lexp}. We differentiate  $C_{ AB}^{[N]}(t)$ with respect to $t$, obtain a sum of $N$ Heisenberg time-derivatives,  and expand each member in powers of $\Delta_l$ to obtain an $N$-fold generalization of \eqnn{begetter}{lambs}. On replacing powers of $\Delta_l$ by powers of
 ${-i\hbar\partial/\partial p_l}$, we obtain
\begin{align}
{ d C_{AB}^{[N]}(t)\over dt} = & {1\over(2\pi\hbar)^N} \int d\bq \int d\bp \no\\
                               & \times \left[ e^{-\beta{\hat H}} {\hat A}\right]_{\overline N}  {\hat L}_{N}\left[ {\hat B}(t)\right]_{N}
\end{align}
where
\begin{align}
 {\hat L}_N =\sum_{l=1}^N& \frac{p_l}{m} \frac{\partial}{\partial q_l} - V(q_l) \frac{2}{\hbar} \sin\!\left(\frac{\ola \partial}{\partial q_l}\frac{\hbar}{2}\frac{\ora \partial}{\partial p_l} \right).
\end{align}
and the arrow notation is as used in \eqn{arrows}.
We can write this expression more compactly in terms of $U_N({\bf q})$ in \eqn{UN} as
\begin{align}
{\hat L}_N ={1\over m}{\bf p}\cdot{\bf \nabla_q} - U_N({\bf q}) \frac{2}{\hbar} \sin\!\left(\frac{\hbar}{2}\,{ {\bf {\ola \nabla}_q}}\cdot{{\bf {\ora \nabla}_p}} \right).
\end{align}
(since all mixed derivatives of $U_N({\bf q})$ are zero).

Following Sec.~IIC, we then truncate the exact Liouvillian at $\hbar^0$ such that
\begin{align}
{\hat L}_N 
 =&{\cal L}_{N}+{\cal O}(\hbar^2)
 \end{align}
 with
\begin{align}
 {\cal L}_{N}=\sum_{l=1}^N \frac{p_l}{m} \frac{\partial}{\partial q_l} - {\partial V(q_l) \over \partial q_l}{\partial  \over \partial p_l}\label{lnq}
\end{align}
The ring-polymer version of LSC-IVR thus approximates the exact dynamics by the classical dynamics of $N$ independent particles, each initiated at a phase-space point $(p_l,q_l)$. The ring-polymer LSC-IVR time-correlation function is
\begin{align}
 C_{AB}^{{\rm W}[N]}(t) = & {1\over(2\pi\hbar)^N} \int d\bq \int d\bp \no\\ 
                          & \times \left[ e^{-\beta{\hat H}} {\hat A}\right]_{\overline N}e^{{\cal L}_{N}t} \left[ {\hat B}(0)\right]_{N}\no\\
                        = & {1\over(2\pi\hbar)^N} \int d\bq \int d\bp \no\\
                          & \times \left[ e^{-\beta{\hat H}} {\hat A}\right]_{\overline N} \left[ {\hat B}(0)\right]_{N}\! ({\bf p}_t,{\bf q}_t)
\end{align}
  where $\left[ {\hat B}(0)\right]_{N}\!
 ({\bf p}_t,{\bf q}_t)$ indicates that this Wigner transform takes its $t=0$ form, but is expressed as a function of the momenta and positions $({\bf p}_t,{\bf q}_t)$ of the $N$ independent particles at time $t$.
  It is easy show (by noting that one can integrate out $N\!-\!1$ of the $p_l$) that
\begin{align}
C_{AB}^{{\rm W}}(t) = \lim_{N\rightarrow\infty}  C_{AB}^{{\rm W}[N]}(t)\label{wolly}
 \end{align}
i.e.\ that the truncation of ${\hat L}_N$ at $\hbar^0$ gives the standard LSC-IVR approximation in the limit $N\to\infty$ (as would be expected, since we have approximated the exact quantum Kubo time-correlation function of \eqnn{monkey}{brolly} by  truncating the quantum Liouvillian at~$\hbar^0$).

\subsection{Normal mode coordinates}
\subsubsection{Definition}
The advantage of ring-polymer coordinates is that we can now transform to sets of global coordinates describing collective motion of the individual coordinates $(p_l,q_l,\Delta_l)$. The choice of global coordinates is not unique. We will find it convenient to use the normal modes of a free ring-polymer,\cite{jeremy,markman} namely
 the linear combinations of $q_l$ that diagonalize $T_N({\bf p},{\bf q})$ of \eqn{tn}. These are simply discrete Fourier transforms, which for odd $N$ (which we will assume, to simplify the algebra\cite{even}), are
\begin{align}
Q_n&=\sum_{l=1}^NT_{ln}q_l, \quad n=0,\pm 1,\dots,\pm (N-1)/2
\end{align}
where
\begin{align}
 T_{ln} = 
 \left\{
 \begin{array}{ll}
  N^{-1/2} & n=0 \\
  \sqrt{2/N} \sin(2\pi ln/N) & n=1,\dots,(N-1)/2 \\
  \sqrt{2/N} \cos(2\pi ln/N) & n=-1,\dots,-(N-1)/2
 \end{array}
 \right.\label{ttt}
\end{align}
and similarly for $P_n$ in terms of $p_l$, and $D_n$ in terms of $\Delta_l$.
The associated normal frequencies take the form
\begin{align}
\omega_n={2\over \beta_N\hbar}\sin{\left(n\pi \over N\right)}\label{norm}
 \end{align}
 such that the ring-polymer expression for $C_{ AB}(0)$ [\eqn{polly}] can be rewritten as
 \begin{align}
C_{ AB}(0) = & \lim_{N\to\infty}{1\over (2\pi\hbar)^N} \int\!d{\bf P}\int\!d{\bf Q} \no\\
             & \times A({\bf Q})B({\bf Q}) e^{-\beta_N R_N({\bf P},{\bf Q})}
\end{align}
where the normal-mode expression for the ring-polymer Hamiltonian $R_N({\bf P},{\bf Q})$ is
 \begin{align}
 R_N({\bf P},{\bf Q}) = \left(\sum_{n=-(N-1)/2}^{(N-1)/2}{P_n^2\over 2m}+{m\over 2}\omega_n^2Q_n^2\right)+U_N({\bf Q})
 \end{align}
and $A({\bf Q})$, $B({\bf Q})$ and $U_N({\bf Q})$ are obtained by making the substitution
\begin{align}
q_l=\sum_{n=-(N-1)/2}^{(N-1)/2}T_{ln}Q_n\label{tback}
\end{align}
into $A({\bf q})$, $B({\bf q})$ and $U_N({\bf q})$ of \eqnt{aaa}{UN}. Note the definition of the sign of $\omega_n$ in \eqn{norm}, which results in somewhat neater expressions later on. Note also that  $R_N({\bf P},{\bf Q})$ will {\em not} be used to generate the dynamics in any of the expressions derived below which, like the dynamics of Sec.~IIIA, will involve $N$ independent particles unconnected by springs.

\subsubsection{Time-correlation functions}

It is straightforward to convert \eqn{disgrace} into normal mode coordinates using the orthogonal transformations in \eqn{tback}, to obtain
\begin{align}
 C_{ AB}^{[N]}(t) = & {1\over(2\pi\hbar)^N} \int d{\bf P} \int d{\bf Q} \no\\
                    & \times \left[ e^{-\beta{\hat H}} {\hat A}\right]_{\overline N}({\bf P},{\bf Q}) \left[ {\hat B}(t)\right]_{N}({\bf P},{\bf Q})\label{wign}
\end{align}
where
\begin{align}
\int d{\bf P}\equiv \prod_{n=-(N-1)/2}^{(N-1)/2}\int_{-\infty}^\infty dP_n
 \end{align}
 and $\int d{\bf Q}$ is similarly defined. The generalized Wigner transforms in \eqn{wign} are obtained using \eqn{tback} to substitute $({\bf P},{\bf Q},{\bf D})$ for $({\bf p},{\bf q},{\bf \Delta})$  in \eqnn{wig1}{wig2}, and thus  contain products of $\exp(iP_nD_n/\hbar)$ in place of 
 $\exp(ip_l\Delta_l/\hbar)$. At $t=0$, one obtains 
\begin{align}
\left[ {\hat B}(0)\right]_{N}({\bf P},{\bf Q}) =B({\bf Q})\label{bpol}
\end{align}
where $B({\bf Q})$ is obtained by substituting ${\bf Q}$ for ${\bf q}$ in $B({\bf q})$ of \eqn{aaa}.

 The (exact) quantum dynamics is described by
\begin{align}
{ d C_{ AB}^{[N]}(t)\over dt} = & {1\over(2\pi\hbar)^N} \int d{\bf P} \int d{\bf Q} \no\\
                                & \times \left[ e^{-\beta{\hat H}} {\hat A}\right]_{\overline N}  {\hat L}_N\left[ {\hat B}(t)\right]_{N}
\end{align}
where the Liouvillian ${\hat L}_N$ is obtained by expressing ${\hat L}_N$ of \eqn{lnq} in terms of normal modes, which gives
\begin{align}
{\hat L}_N ={1\over m}{\bf P}\cdot{\bf \nabla_Q} - U_N({\bf Q}) \frac{2}{\hbar} \sin\!\left(\frac{\hbar}{2}\, {{\bf {\ola \nabla}_Q}}\cdot{{\bf {\ora \nabla}_P}} \right).\label{norty}
\end{align}
in which $U_N({\bf Q})$ is obtained by substituting ${\bf Q}$ for ${\bf q}$ in $U_N({\bf q})$ of \eqn{UN}.

As in Sec.~IIIA, the LSC-IVR dynamics is obtained by truncating ${\hat L}_N$ at $\hbar^0$ to give
\begin{align}
{\cal L}_{N}=\sum_{n=-(N-1)/2}^{(N-1)/2}{{ P}_n\over m}{\partial \over \partial { Q}_n}
  - {\partial { U}_N({\bf  Q})\over\partial { Q}_n}{\partial \over\partial { P}_n}\label{norml}
\end{align}
after which one obtains  $C_{ AB}^{{\rm W}[N]}(t)$  in terms of normal modes,
which gives the (standard) LSC-IVR result in the limit \ninc according to
\eqn{wolly}. Hence all we have done in \eqnt{wign}{norml} is to re-express the
results of Sec.~IIIA in terms of normal mode coordinates. The advantages of
doing this will become clear shortly.

 \subsection{Matsubara modes}
 
We now consider the $M$ lowest frequency ring-polymer normal modes in the limit \ninc such that
 $M\ll N$. The frequencies $\omega_n$ tend to the values
\begin{align}
{\widetilde\omega}_n=\lim_{N\rightarrow\infty}\omega_n={2n\pi \over \beta\hbar},\quad |n| \le (M-1)/2\label{matty}
\end{align}
which are often referred to as the `Matsubara frequencies',\cite{matsubara} and so we will refer to these $M$ modes in the limit \ninf as the `Matsubara modes'. The Matsubara modes have the special property that any superposition of them produces a distribution of the coordinates $q_l$ which is a smooth and differentiable function of imaginary time $\tau$, such that 
\begin{align}
q_l=q(\tau),\quad \tau=\beta_N\hbar\, l,\quad\quad l=1,\dots,N
\end{align}
(see Appendix A).
Hence distributions made up of superpositions of the Matsubara modes resemble the sketch in Fig.~2.
We will often write the Matsubara modes using the notation
\begin{align}
{\widetilde Q}_n=\lim_{N\to\infty}{Q_n\over\sqrt{N}},\quad n=0,\pm 1,\dots,\pm (M-1)/2
\end{align}
(and similarly for $\wt P_n$, $\wt D_n$). The extra factor of $N^{-1/2}$ ensures that  $\wt Q_n$ scales as $N^0$ and converges in the limit $N\to\infty$; e.g.\  $\wt Q_0$ is the centroid (centre of mass) of the smooth distribution $q(\tau)$. We will refer to the other $N-M$ normal modes as the `non-Matsubara modes'. In general, these modes give rise to jagged (i.e.\ non-smooth, non-differentiable with respect to $\tau$) distributions of $q_l$ (see Fig.~2).\cite{arti} 

\begin{figure}[t]
    \begin{center}
        \centering
        \includegraphics[scale=0.60]{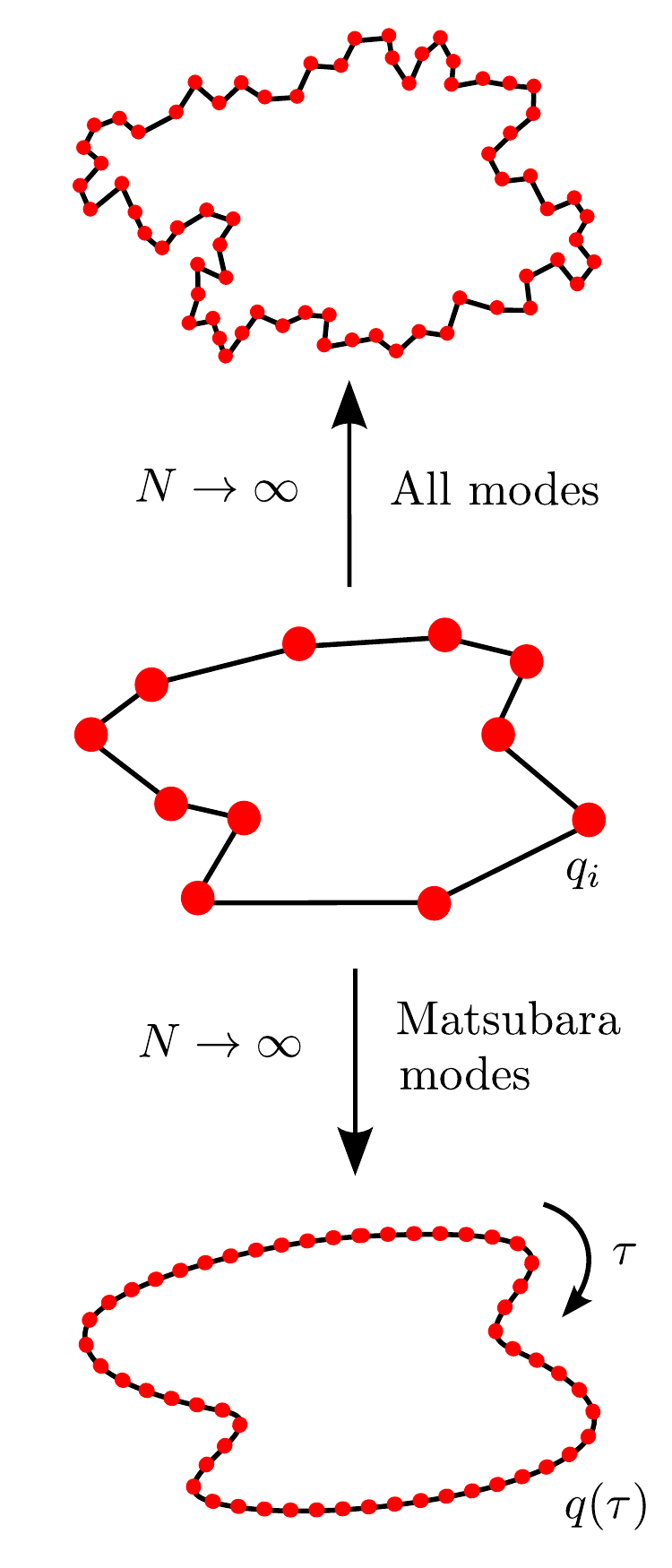}
        \caption{Schematic diagram showing that superpositions of Matsubara
            modes give distributions of path-integral coordinates $q_l$ which
            are smooth, differentiable functions of imaginary time $\tau$.
            Inclusion of non-Matsubara modes gives jagged distributions.}
        \label{figure2}
    \end{center}
\end{figure}
 
Matsubara modes have a long history\cite{ceperley,charu,doll,char_cep} in path-integral descriptions of equilibrium properties, since they give rise to an alternative ring-polymer expression for
$C_{ AB}(0)$. If we define 
\begin{align}
 C_{ AB}^{[M]}(0) ={\alpha_M\over 2\pi\hbar}\int d{\bf \widetilde P}& \int d{\bf \widetilde Q}\ A({\bf \widetilde Q})B({\bf \widetilde Q})e^{-\beta \wt R_M
 ({\bf \wt P},{\bf \wt Q})}\label{matpol}
\end{align}
with
 \begin{align}
 \wt R_M({\bf \wt P},{\bf \wt Q}) =& \left(\sum_{n=-(M-1)/2}^{(M-1)/2}{{\wt P}_n^2\over 2m}+{m\over 2}\wt\omega_n^2\wt Q_n^2\right)+\wt U_M({\bf \wt Q})\\
{\widetilde U}_M({\bf {\widetilde Q}})=&\lim_{N\to\infty}{1\over N}\sum_{l=1}^N V\left(\sum_{n=-(M-1)/2}^{(M-1)/2}{T}_{ln}\sqrt{N}{\widetilde Q}_n\right)\\
\alpha_M=&\hbar^{(1-M)}\left[(M-1)/ 2\right]!^2
\end{align}
then
\begin{align}
C_{ AB}(0) = \lim_{\substack{M\to\infty\\M\ll N}}C_{ AB}^{[M]}(0)\label{exactus}
\end{align}
where this limit indicates that $M$ is allowed to tend to infinity, subject to
the condition that it is always much smaller than $N$, such that the ${\bf \wt
Q}$ remain Matsubara modes. In practice, a good approximation to the exact
result is reached once $\wt \omega_{(M-1)/2}$ exceeds the highest frequency in
the potential~$V(q)$. \Eqn{matpol} is less often used nowadays to compute static
properties, because the convergence with respect to $M$ is typically slower than
the convergence of \eqn{polly} with respect to $N$.\cite{char_cep}

However, \eqn{matpol} tells us something interesting: The Boltzmann factor ensures that only smooth distributions of $({\bf p},{\bf q})$ survive in $C_{ AB}(t)$ at $t=0$; but at $t>0$,  the force terms in ${\cal L}_{N}$ [\eqn{norml}] will, in general, mix in an increasing proportion of non-smooth, non-Matsubara modes, so that the distributions of $({\bf p},{\bf q})$ become increasingly jagged as time evolves. The rate at which this mixing occurs depends on the anharmonicity of the potential $V(q)$.  In the special case that $V(q)$ is harmonic, there is no coupling between different normal modes, so the distributions in $({\bf p},{\bf q})$ remain smooth for all time.  In other words, smooth distributions in $({\bf p},{\bf q})$ are found in two of the limits
(zero-time and harmonic) in which the LSC-IVR is known to be exact. 

\section{Matsubara dynamics}

\subsection{Definition}

The results of Sec.~IIIC suggest that there may be a  connection between smoothness in imaginary time and classical dynamics. We now investigate what happens if we constrain an initially smooth function of phase space coordinates $({\bf p},{\bf q})$ to remain smooth for all (real) times $t>0$. We take the (exact) quantum Liouvillian ${\hat L}_N$, and instead of truncating at $\hbar^0 $ as in \eqn{norml} (which gives the LSC-IVR), we retain {\em all} powers of $\hbar^2$, take the \ninf limit, and split  ${\hat L}_N$ into \begin{align}
\lim_{N\to\infty}{\hat L}_N = {\cal L}_{M} +\lim_{N\to\infty} {\hat L}_{\rm error}(N,M) \label{things}
\end{align}
where the `Matsubara Liouvillian' 
\begin{align}
 {\cal L}_{M} = & \lim_{N\to\infty}\sum_{n=-(M-1)/2}^{(M-1)/2}{P_n\over m}{\partial \over \partial Q_n} \no\\
  & -  U_N({\bf Q}) \frac{2}{\hbar} \sin\!\left(\sum_{n=-(M-1)/2}^{(M-1)/2}   \frac{\hbar}{2}{\ola \partial \over \partial Q_n}{\ora \partial \over \partial P_n} \right)\label{matlouf}
\end{align}
contains all terms in which the derivatives involve {\em only} the Matsubara modes, and
${\hat L}_{\rm error}(N,M)$ contains the rest of the terms (given in Appendix B). We then discard ${\hat L}_{\rm error}(N,M)$, approximating ${\hat L}_N$ by ${\cal L}_{M}$. We will refer to the (approximate) dynamics generated by ${\cal L}_{M}$ as `Matsubara dynamics'. By construction, Matsubara dynamics ensures that a distribution of $({\bf p},{\bf q})$ which is a smooth and differentiable function of $\tau$ at $t=0$ will remain so for all  $t>0$.

The time-correlation function corresponding to Matsubara dynamics is
\begin{align}
 C_{AB}^{[M]}(t) = & \lim_{N\rightarrow\infty} {1\over(2\pi\hbar)^N}\int d{\bf P} \int d{\bf Q} \no\\
                   & \times \left[ e^{-\beta{\hat H}} {\hat A}\right]_{\overline N}e^{{\cal L}_{M}t} \left[ {\hat B}(0)\right]_{ N}\label{blobby}
\end{align}
We can obtain an explicit form for $C_{ AB}^{[M]}(t)$ by 
taking the same limit as in \eqn{exactus},  allowing $M$ to tend to infinity, 
 subject to $M\ll N$, which gives (see Appendix C)
\begin{align}
C_{ AB}^{\rm Mats}(t)=\lim_{\substack{M\to\infty\\M\ll N}} C_{ AB}^{[M]}(t)
 \end{align}
where
\begin{align}
 C_{ AB}^{[M]}(t) ={\alpha_M\over 2\pi\hbar}\int d{\bf \widetilde P}& \int d{\bf \widetilde Q}\ A({\bf \widetilde Q})e^{-\beta[{\wt H}_M({\bf \wt P},{\bf \wt Q})-i\theta_M({\bf \widetilde P},{\bf \widetilde Q})]}\no\\
&\times e^{{\cal L}_{M}t}B({\bf \widetilde Q})\label{mattsc}
\end{align}
in which the Matsubara Hamiltonian is
\begin{align}
{\wt H}_M({\bf \wt P},{\bf \wt Q})={{\bf \widetilde P}^2\over 2m}+{\widetilde U}_M({\bf \widetilde Q})
\end{align}
and the phase factor is 
\begin{align}
\theta_M({\bf \widetilde P},{\bf \widetilde Q})=\sum_{n=-(M-1)/2}^{(M-1)/2}{\widetilde P}_n{\widetilde\omega}_n{\widetilde Q}_{-n}\label{thet}
\end{align}
with $\alpha_M$, $\widetilde\omega_n$, ${\bf \widetilde P}$ and ${\bf \widetilde Q}$   defined  in Sec.~IIIC. Note that, in deriving these equations (in Appendix C), we have not proved that $C_{ AB}^{[M]}(t)$ converges with $M$ for $t>0$ (only that the form of \eqnt{mattsc}{thet} converges with $M$). We test this convergence numerically in Sec.~V.

Thus when the exact dynamics is approximated by Matsubara dynamics, the quantum Boltzmann distribution takes the simple form of a classical Boltzmann distribution multiplied by a phase factor. At $t=0$, one may analytically continue the phase factor (by making $P_n\to P_n-im\omega_nQ_{-n}$) to recover the ring-polymer distribution in \eqn{matpol}. However, it is not known whether this analytic continuation is valid at $t>0$ (except for the special case of the harmonic oscillator), and hence the most general form of quantum Boltzmann distribution (in the space of Matsubara modes) is the one given in \eqn{mattsc}.

\subsection{Matsubara dynamics is classical}

We now rewrite ${\cal L}_{M}$ in terms of $({\bf \wt P},{\bf \wt Q})$, to make explicit its dependence on $N$, and we also assume that $M$ is sufficiently large that \eqn{mattsc} holds, allowing us to replace~${U}_N({\bf {Q}})/N$ by ${\widetilde U}_M({\bf {\widetilde Q}})$. This gives
\begin{align}
{\cal L}_{M}  =\lim_{N\to\infty}{1\over m}{\bf {\widetilde P}}\cdot{\bf \nabla_{\widetilde Q}} - {\widetilde U}_M({\bf {\widetilde Q}}) \frac{2N}{\hbar} \sin\!\left(\frac{\hbar}{2N}\ {{\bf {\ola \nabla}_{\widetilde Q}}}\cdot{{\bf {\ora \nabla}_{\widetilde P}}} \right).\label{hn}
\end{align}
In other words,  the Moyal series in Matsubara space\cite{arti2} is an expansion in terms of $(\hbar/N)^2$, rather than  $\hbar^2$. Now, it is well known\cite{heller} that the smallness of $\hbar$ cannot in general be used to justify truncating the (standard LSC-IVR) Moyal series of \eqn{lexp} at $\hbar^0$, since at least one of the Wigner transforms in the time-correlation function [\eqn{cabw}]  contains derivatives that scale as $\hbar^{-1}$.  However, it is easy to show that the derivatives of all terms in the integral in \eqn{mattsc} scale as $N^0$. As a result, it follows that all derivatives higher than first order in ${\cal L}_{M}$ vanish in the limit \ninc with the result that
\begin{align}
{\cal L}_{M}=\sum_{n=-(M-1)/2}^{(M-1)/2}{{\widetilde P}_n\over m}{\partial \over \partial {\widetilde Q}_n}
  - {\partial {\widetilde U}_M({\bf \widetilde Q})\over\partial {\widetilde Q}_n}{\partial \over\partial {\widetilde P}_n}
\end{align}
In other words, Matsubara dynamics is {\em classical}.

This is a surprising result, which needs to be interpreted with caution.  It does not mean that the dependence of $B({\bf Q})$ on the Matsubara modes evolves classically in the exact quantum dynamics, since the exact Liouvillian ${\hat L}_N$
contains derivative terms that couple the Matsubara modes with the non-Matsubara modes (for which the higher-order derivatives cannot be neglected): it means that the dynamics of the Matsubara modes becomes classical when they are {\em decoupled} from the non-Matsubara modes.

One way to understand the origin of the $\hbar/N$ in \eqn{hn} is to note that the Fourier transform between ${\wt P}_n$ and ${\wt D}_n$ (in the Wigner transforms of \eqnn{wig1}{wig2}) is $\exp(iN{\wt P}_n{\wt D}_n/\hbar)$. Hence the effective Planck's constant associated with motion in the Matsubara coordinates tends to zero in the limit $N\to\infty$.  Note that the dependence of the Boltzmann distribution on the non-Matsubara modes is more complicated than that of \eqn{mattsc}, and contains powers of~$(\hbar/N)^{-1}$ which cancel out the powers of
$(\hbar/N)$ in ${\hat L}_N$ (which must obviously happen, since we know that the exact dynamics is not in general classical).

Matsubara dynamics thus has many features in common with LSC-IVR: it is exact in the $t=0$ limit (when all distributions of $({\bf p},{\bf q})$ are smooth superpositions of Matsubara modes), in the harmonic limit (where the dynamics of the Matsubara modes is decoupled from that of the non-Matsubara modes), and in the classical limit (since setting $M=0$ in \eqn{mattsc} gives the classical time-correlation function); and it neglects all terms  ${\cal O}(\hbar^2)$ in the (exact) quantum Liouvillian. However, Matsubara dynamics differs from LSC-IVR in that it also neglects the terms ${\cal O}(\hbar^0)$ that contain derivatives in the non-Matsubara modes. One can thus regard Matsubara dynamics as a filtered version of LSC-IVR, in which the parts of the dynamics that cause the smooth distributions of $({\bf p},{\bf q})$ to become jagged have been removed.\cite{der2}  

\subsection{Conservation of the quantum Boltzmann distribution}

Confining the dynamics to the space of Matsubara modes has a major effect on the symmetry of the Hamiltonian. The LSC-IVR Hamiltonian ${H}_N({\bf  P},{\bf  Q})$ is simply the classical Hamiltonian of $N$ independent particles, and  is thus symmetric with respect to any permutation of the phase space coordinates [e.g.\ $(p_1,q_1)\leftrightarrow(p_3,q_3)$]. On restricting the dynamics to the Matsubara modes, most of these symmetries are lost (since individual permutations would destroy the smoothness of the distributions of $({\bf p},{\bf q})$). However, one operation which is retained\cite{reflect} is symmetry with respect to {\em cyclic} permutation of the coordinates, which, on restricting the dynamics to Matsubara space, becomes a continuous, differentiable symmetry, namely invariance with respect to translation in imaginary time:
\begin{align}
{d{\wt H}_M({\bf \wt P},{\bf \wt Q})\over d\tau}=0\label{imtrans}
\end{align}
(see Appendix A). It thus follows from Noether's theorem,\cite{goldstein} that
\begin{align}
{d \wt \Lambda_M({\bf \wt P},{\bf \wt Q}) \over d\tau}={d\over dt}\left(  \sum_{n=-(M-1)/2}^{(M-1)/2}{\wt P}_n{d {\wt Q}_n\over d\tau}   \right)=0
\end{align}
where $\Lambda_M({\bf \wt P},{\bf \wt Q})$ is the Matsubara Lagrangian. In other words, in Matsubara dynamics, there exists a constant of the motion (in addition to the total energy) which is given by the term in brackets above.

In Appendix A, it is shown that the phase $\theta_M({\bf \wt P},{\bf \wt Q})$ in the quantum Boltzmann distribution [\eqnt{mattsc}{thet}] can be written
\begin{align}
\theta_M({\bf \wt P},{\bf \wt Q})=-\!\!\!\!\!\!\!\sum_{n=-(M-1)/2}^{(M-1)/2}{\widetilde P}_n{d {\widetilde Q}_n\over d\tau}\label{thder}
\end{align}
and is thus the constant of the motion associated with the invariance of ${\wt H}_M({\bf \wt P},{\bf \wt Q})$ to imaginary time-translation.
Since ${\wt H}_M({\bf \wt P},{\bf \wt Q})$ is of course also a constant of the motion, it follows that Matsubara dynamics {\em conserves the quantum Boltzmann distribution}.

As a result, Matsubara dynamics satisfies  the detailed balance relation
\begin{align}
 C_{ AB}^{[M]}(t) =C_{ BA}^{[M]}(-t) 
 \end{align}
and gives expectation values
 \begin{align}
 \big<{\hat B} \big>^{[M]}(t) = & {\alpha_M\over 2\pi\hbar} \int d{\bf \widetilde P} \int d{\bf \widetilde Q} \no\\
 & \times e^{-\beta[{\wt H}_M({\bf \widetilde P},{\bf \widetilde Q})-i\theta_M({\bf \widetilde P},{\bf \widetilde Q})]}B({\bf \widetilde Q}_t)\no\\
 = & {\alpha_M\over 2\pi\hbar} \int d{\bf \widetilde P}_t \int d{\bf \widetilde Q}_t \no\\ 
 & \times e^{-\beta[{\wt H}_M({\bf \widetilde P}_t,{\bf \widetilde Q}_t)-i\theta_M({\bf \widetilde P}_t,{\bf \widetilde Q}_t)]}B({\bf \widetilde Q}_t)\no\\
  = & {\alpha_M\over 2\pi\hbar} \int d{\bf \widetilde P}_t \int d{\bf \widetilde Q}_t \no\\
  & \times e^{-\beta{\wt R}_M({\bf \widetilde P}_t,{\bf \widetilde Q}_t)}B({\bf \widetilde Q}_t)\no\\
  = & \big<{\hat B} \big>^{[M]}(0)  \label{expy}
 \end{align}
which are independent of time (and equal to the exact quantum result in the limit $M\to\infty$; see \eqn{exactus}). Note that the  step between the second and third lines follows from analytic continuation ($P_n\to P_n-im\omega_nQ_{-n}$). 

We thus have the surprising result that a purely classical dynamics (Matsubara dynamics) which uses the smoothed Hamiltonian that arises naturally when the space is restricted to Matsubara modes, conserves the quantum Boltzmann distribution. At first sight this may appear counter-intuitive. For example, it is clear that the classical dynamics will not respect zero-point energy constraints, nor will it be capable of tunnelling. However, it is the phase $\theta_M({\bf \widetilde P},{\bf \widetilde Q})$ which converts what would otherwise be a classical Boltzmann distribution in an extended phase-space into a quantum Boltzmann distribution, and the phase is conserved. 

\subsection{Generalizations}

The derivations above can easily be generalized to systems with any number of dimensions. For a system whose classical Hamiltonian resembles
\eqn{classf}, there are $F\times M$ Matsubara modes,  one set of $M$ modes in each  dimension. All the steps in Secs.~III and IV.A-C are then the same, except that, with $F$ dimensions instead of one, there is now a sum of $F$ phase terms, each resembling $\theta_M({\bf \widetilde P},{\bf \widetilde Q})$. Noether's theorem  shows that the sum of these terms and hence the quantum Boltzmann distribution is conserved.

We emphasise that the derivations above were carried out for operators ${\hat A}$ and ${\hat B}$ in $C_{ AB}(t)$ which are {\em general} functions of the coordinate operators ${\hat q}$. Matsubara dynamics is therefore not limited to correlation functions involving linear operators of position.  The derivations can also be repeated, with minor modifications in the algebra, for the case that ${\hat A}$ and ${\hat B}$ are general functions of the momentum operator (which results in functions of ${\bf \wt P}$ appearing in the generalised Wigner transforms).

\section{Numerical tests of the efficacy of Matsubara dynamics}

\begin{figure}[t]
    \begin{center}
        \centering
        \includegraphics[scale=0.60]{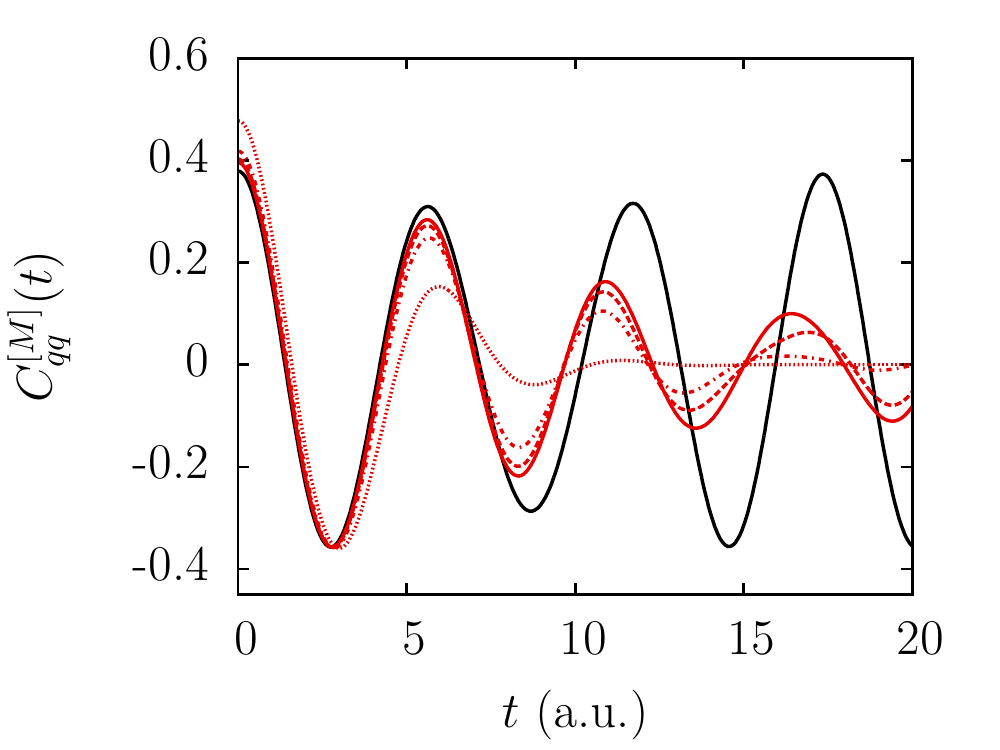}
        \caption{Convergence with respect to number of modes $M$ of the
            Matsubara position auto-correlation function $C_{ qq}^{[M]}(t)$,
            calculated for the quartic potential of \eqn{quart}, at a reciprocal
            temperature of $\beta=2$ a.u. The red lines correspond to $M=1$
            (dots), 3 (chains), 5 (dashes) and 7 (solid). The solid black line
            is the exact quantum result.}
        \label{figure3}
    \end{center}
\end{figure}

\begin{figure}[t]
    \begin{center}
        \centering
        \includegraphics[scale=0.60]{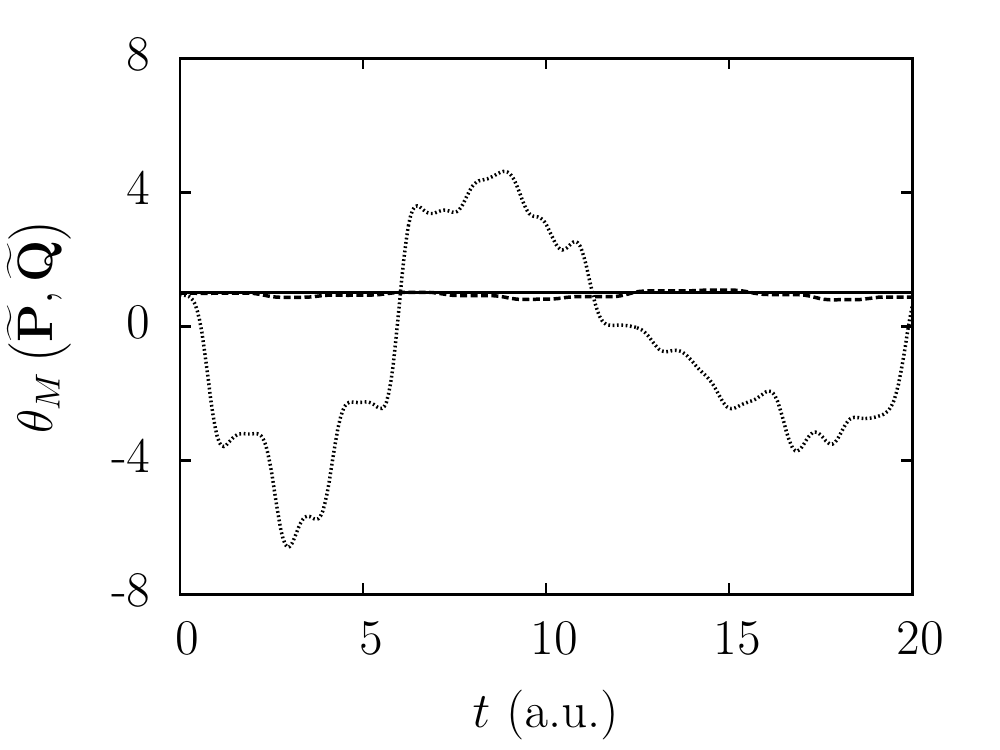}
        \caption{Evolution of the phase $\theta_M({\bf \widetilde P},{\bf
           \widetilde Q})$ along a single classical trajectory on the quartic
           potential, with $M=5$, and $N=5$ (dots), 9 (dashes) and $\infty$ (solid
           line). The latter corresponds to Matsubara dynamics in which the phase
           is conserved.}
        \label{figure4}
    \end{center}
\end{figure}

So far we have made no attempt to justify the use of Matsubara dynamics, beyond pointing out that it is exact in all the limits in which  LSC-IVR is exact, but that, unlike LSC-IVR, it also conserves the quantum Boltzmann distribution. Here we investigate whether Matsubara dynamics converges with respect to the number of modes $M$, and make numerical comparisons with the LSC-IVR, CMD and RPMD methods.

The presence of the phase $\theta_M({\bf \widetilde P},{\bf \widetilde Q})$ in the Boltzmann distribution [\eqn{mattsc}] means that Matsubara dynamics suffers from the sign problem, and thus cannot be used as a practical method.  However, we were able to evaluate $C_{ qq}^{[M]}(t)$ (i.e.\ $C_{ AB}^{[M]}(t)$ of \eqn{mattsc} with ${\hat A}=\hat q$, ${\hat B}=\hat q$) for some
 one-dimensional model systems. For consistency with previous work,\cite{craig1,rossky} we considered the quartic potential
\begin{align}
 V(q) = \frac{1}{4}q^4\label{quart}
\end{align}
and the weakly anharmonic potential
\begin{align}
 V(q) = \frac{1}{2}q^2 + \frac{1}{10}q^3 + \frac{1}{100}q^4\label{weak}
\end{align}
where atomic units are used with $m=1$. Calculations using potentials with intermediate levels of anharmonicity were found to give similar results (and are not shown here).

\begin{figure}[t]
    \begin{center}
        \centering
        \includegraphics[scale=0.60]{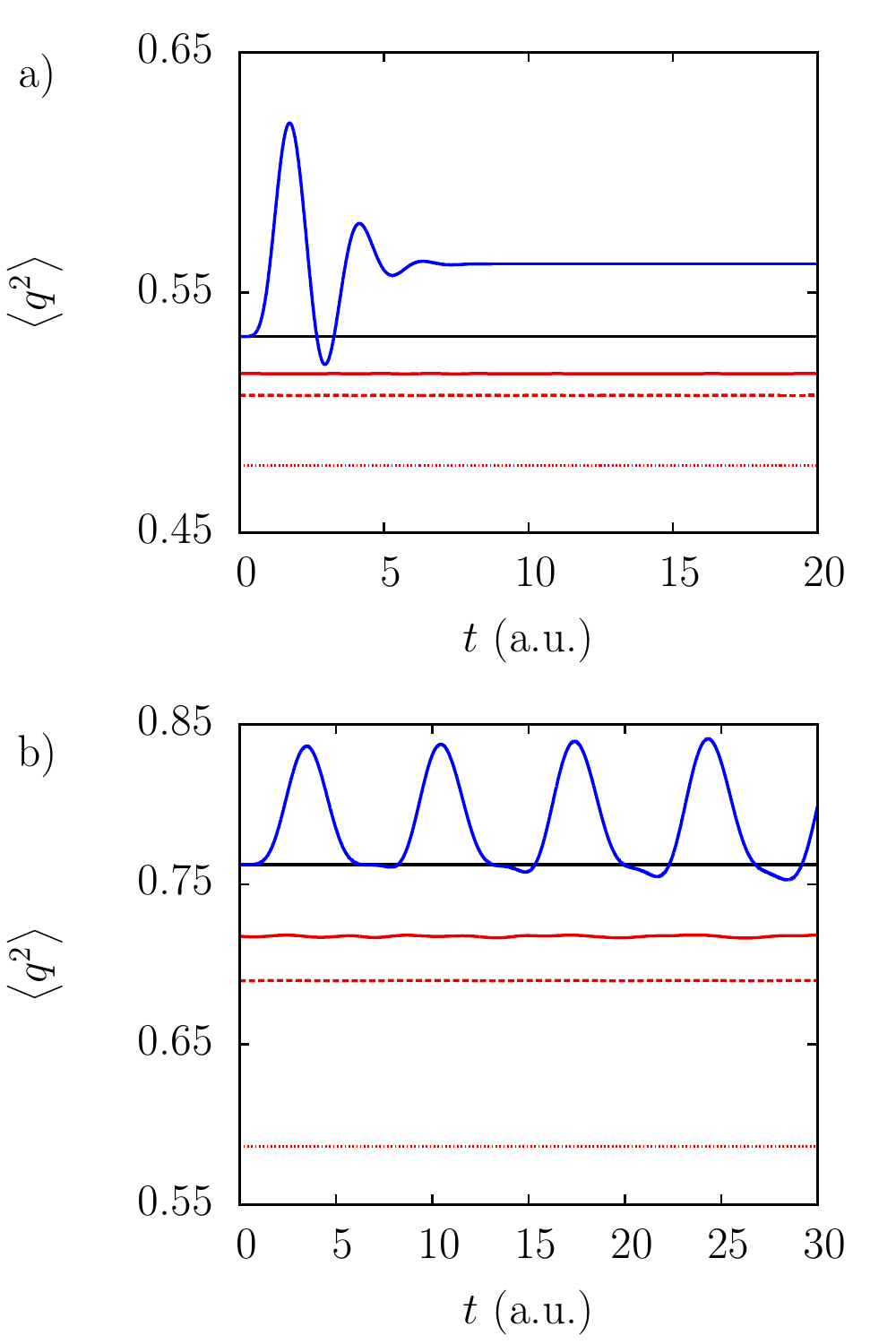}
        \caption{Time-dependence of the thermal expectation value
            $\big<q^2\big>(t)$ for the quartic potential at $\beta=2$, computed
            using LSC-IVR (blue), and  Matsubara dynamics (red: $M=1$ (dots), 3
            (dashes), 5 (solid)), and compared with the exact quantum result
            (black).}
        \label{figure5}
    \end{center}
\end{figure}

\begin{figure}[t]
    \begin{center}
        \centering
        \includegraphics[scale=0.60]{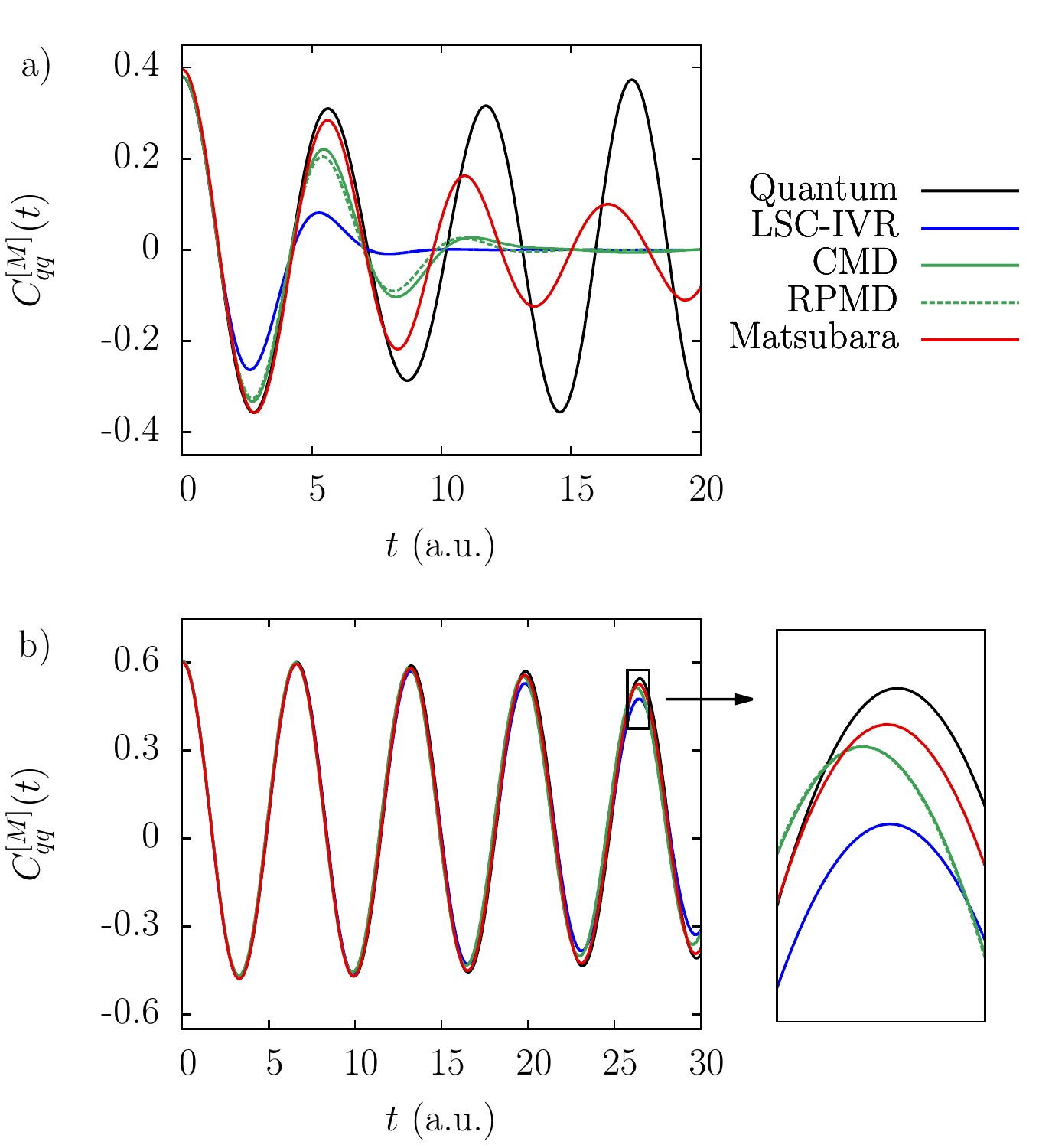}
        \caption{Comparisons of position-autocorrelation functions computed
            using different levels of theory, for (a) the quartic potential and
            (b) the weakly anharmonic potential of \eqn{weak}. The Matsubara
            results were obtained using $M=7$ (quartic) and $M=5$ (weakly
            anharmonic).}
        \label{figure6}
    \end{center}
\end{figure}

Figure 3 shows  $C_{ qq}^{[M]}(t)$ for the quartic potential, at an inverse temperature of $\beta=2$ a.u., for various values of $M$. These results were obtained by propagating classical trajectories using the Matsubara potential ${\widetilde U}_M({\bf \widetilde Q})$ to generate the forces, subject to the Anderson thermostat\cite{daan} (according to which each $\wt P_n$ was reassigned to a value drawn at random from the classical Boltzmann distribution every 2 atomic time units);   ${\widetilde U}_M({\bf \widetilde Q})$ was computed by taking the \ninf limit analytically,  as described in the supplemental material.\cite{suppl} A total of $10^{11}$ Monte Carlo points was found necessary to converge $C_{qq}^{[M]}(t)$.  Extending these calculations 
beyond $M=7$ was prohibitively expensive, and the final few $M$ were particularly difficult to converge (since $\theta_M({\bf \widetilde P},{\bf \widetilde Q})$ becomes increasingly oscillatory as  $\wt \omega_n$ increases). Nevertheless, the results in Fig.~3 are sufficient to show that $C_{ qq}^{[M]}(t)$ converges with respect to $M$, although the convergence appears to become slower as $t$ increases.  For the weakly anharmonic potential,  convergence to within graphical accuracy was obtained using $M=5$ for  $\beta=2$ a.u. 

We also confirmed numerically that Matsubara dynamics conserves the quantum Boltzmann distribution. Figure 4 shows the phase $\theta_M({\bf \widetilde P},{\bf \widetilde Q})$ as a function of time along a Matsubara trajectory. When a coarse number of polymer beads ($N=5$) is used, such that the $M$ lowest-frequency modes are a poor approximation to the Matsubara modes,  the phase is not conserved; however, as $N$ is increased, the variation of the phase along the trajectory flattens, becoming completely time-independent in the limit $N\to\infty$.   Figure 5 plots the expectation value $\big<q^2\big>^{[M]}(t)$, which is found to be time-independent as expected from \eqn{expy}.

Figure 6 compares the Matsubara correlation functions $C_{ qq}^{[M]}(t)$ for both potentials with exact quantum, LSC-IVR, CMD and RPMD results. The quartic potential  at $\beta=2$ (panel~a) is a severe test for which any method that neglects real-time coherence fails after a single recurrence. Nevertheless, we see that Matsubara dynamics gives a much better treatment than LSC-IVR, reproducing almost perfectly the first recurrence at 6 a.u., and damping to zero more slowly.\cite{sc-ivr} The Matsubara result is also better than both the CMD and RPMD results. The same trends are found for the weakly anharmonic potential (Fig.~6, panel b), and were also found for the potentials with intermediate anharmonicity (not shown).

\section{Conclusions}

We have found that a single change in the derivation of LSC-IVR dynamics gives rise to a classical dynamics (`Matsubara dynamics') which preserves the quantum Boltzmann distribution. This change involves no explicit truncation in powers of $\hbar^2$, but instead a decoupling of a subspace of ring-polymer normal modes (the Matsubara modes) from the other modes. The dynamics in this restricted space is found to be purely classical and to ensure that smooth distributions of phase-space points (as a function of imaginary time), which are present in the Boltzmann distribution at time $t=0$, remain smooth at all later times.
 The LSC-IVR dynamics, by contrast, includes all the modes, which has the effect of breaking up these smooth distributions, and thus failing to preserve the quantum Boltzmann distribution. Numerical tests show that Matsubara dynamics gives consistently better agreement than LSC-IVR with the exact quantum time-correlation functions.
 
These results suggest that Matsubara dynamics is a better way than LSC-IVR, at least in principle, to account for the classical mechanics in quantum time-correlation functions. We suspect that Matsubara dynamics may be equivalent to expanding the time-dependence of the  quantum time-correlation function in powers of $\hbar^2$ and truncating it at $\hbar^0$; this is in contrast to LSC-IVR, in which one truncates the quantum Liouvillian\cite{high} at $\hbar^0$. However, further work will be needed to prove or disprove this conjecture.

Matsubara dynamics is far too expensive to be useful as a practical method. However,  it is probably a good starting point from which to make further approximations in order to develop such methods.  The numerical tests reported here show that  Matsubara dynamics gives consistently better results than both CMD and RPMD, suggesting that these popular methods may be approximations to Matsubara dynamics.

\begin{acknowledgments}
TJHH, MJW and SCA acknowledge funding from the UK Science and Engineering Research Council. AM acknowledges the European Lifelong Learning Programme (LLP) for an Erasmus student placement scholarship. TJHH also acknowledges a Research Fellowship from Jesus College, Cambridge and helpful discussions with Dr Adam Harper.
\end{acknowledgments}

\appendix

  \section{Differentiability with respect to imaginary time}
 
 A distribution of ring-polymer coordinates $q_l,\, l=1,\dots,N$, can be written as a smooth and differentiable function of the imaginary time $\tau$ ($0\le\tau<\beta\hbar$) if the limit 
  \begin{align}
{d q(\tau)\over d\tau}=\lim_{N\to\infty} {q_{l+1}-q_{l-1}\over 2\beta_N\hbar},\quad \tau=l\hbar\beta_N
 \end{align}
exists, i.e.\ if
   \begin{align}
\lim_{N\to\infty} q_{l+1}-q_{l-1}\sim N^{-1}\label{limn}
 \end{align}
For a distribution formed by superposing {\em only} the Matsubara modes,  we can use trigonometric identities and the definitions in Sec.~III to write
  \begin{align}
& q_{l+1}-q_{l-1} \no\\
& = 2\sqrt{2}\sum_{n=1}^{(M-1)/2}\left[\cos\left({2\pi nl\over N}\right){\wt
 Q}_n-\sin\left({2\pi nl\over N}\right){\wt Q}_{-n} \right] \no\\
& \ \ \ \times \sin\left({2\pi n\over N}\right) 
 \end{align}
Since $n\ll N$, the sine function on the right ensures that \eqn{limn} is satisfied; also, repetition of this procedure shows that higher-order differences of order $\lambda$ scale as $N^{-\lambda}$. Hence a distribution $q_l$ formed from a superposition of Matsubara modes is a smooth and differentiable function of $\tau$. The same is true for distributions in $p_l$ and $\Delta_l$. 

 To prove that the Matsubara Hamiltonian is invariant under imaginary-time translation [\eqn{imtrans}], 
 we first differentiate the Matsubara potential $U_M ({\bf \wt Q})$ with respect to $\tau$, which gives 
 \begin{align}
{d \wt U_M ({\bf \wt Q})\over d\tau}=\lim_{N\to\infty}{\wt {\cal P}_{1}\,
\wt U_M({\bf q})- \wt U_M({\bf q})\over \beta_N\hbar}
 \end{align}
where 
\begin{align}
\wt U_M({\bf q})=\sum_{l=1}^NV\left(\sum_{m=1}^N\ \sum_{n=-(M-1)/2}^{(M-1)/2} T_{ln}T_{mn}q_{m}  \right)
 \end{align}
 and $ {\cal P}_{1} $
  represents  a cyclic permutation of the coordinates $q_m\to q_{m+1}$, such that
\begin{align}
{\cal P}_{1}\,
\wt U_M({\bf q})=\sum_{l=1}^NV\left(\sum_{m=1}^N\ \sum_{n=-(M-1)/2}^{(M-1)/2} T_{ln}T_{(m-1)\,n}q_{m}  \right)\label{pum}
 \end{align}
 We then rearrange the sum over $n$ in \eqn{pum} into
 \begin{align}
T_{l0}T_{(m-1)\,0}+ \sum_{n=1}^{(M-1)/2} \left[ T_{ln}T_{(m-1)\,n} + T_{l\,-n}T_{(m-1)\,-n} \right]
 \end{align} 
and use trigonometric identities to show that
\begin{align}
& T_{ln}T_{(m-1)\,n}+T_{l\,-n}T_{(m-1)\,-n} \no\\
& =T_{(l+1)\,n}T_{mn}+T_{(l+1)\,-n}T_{m\,-n}
 \end{align}
Re-ordering the sum over $l$ and using the property that $T_{l0}=N^{-1/2}$ gives
\begin{align}
 {\cal P}_{1}\,
\wt U_M({\bf q})=\wt U_M({\bf q})
\end{align}
 which proves that 
\begin{align}
{d \wt U_M ({\bf \wt Q})\over d\tau}=0 \end{align}
The same line of argument can be applied to the kinetic energy ${\bf \wt P}^2/2m$, thus proving \eqn{imtrans}.
 
To obtain the derivative of $\wt Q_n$ with respect to $\tau$ (needed to prove \eqn{thder}), we write
\begin{align}
{d\wt Q_n\over d\tau}=&\lim_{N\to\infty}{1\over\sqrt{N}}\sum_{l=1}^NT_{ln}{q_{l+1}-q_{l-1}\over 2\beta_N\hbar}\no\\
=&\lim_{N\to\infty}{1\over\sqrt{N}}\sum_{l=1}^N{[T_{(l-1)\,n}-T_{(l+1)\,n}]q_{l}\over 2\beta_N\hbar}
\end{align}
and use trigonometric identities to obtain
\begin{align}
T_{l+1\,n}-T_{l-1\,n}=2T_{l\, -n}\sin(2n\pi/N)
\end{align}
Since $n\ll N$, it follows that
\begin{align}
{d\wt Q_n\over d\tau}=-{\wt \omega}_n\wt Q_{-n}
\end{align}
where $\wt\omega_n$ is the Matsubara frequency defined in \eqn{matty}.

\section{Error term for Matsubara Liouvillian}

The error term ${\hat L}_{\rm error}(N,M)$  of \eqn{things}
 is the difference ${\hat L}_N-{\cal L}_M$ between the exact quantum Liouvillian and the Matsubara Liouvillian. Using \eqnn{norty}{matlouf} and the trigonometric identity
\begin{align}
\sin(a+b)-\sin a\equiv  2\sin \left({b\over 2}\right)\cos\left(a+{b\over 2}\right)
\end{align}
we can write
\begin{align} 
{\hat L}_{\rm error}(N,M) = & \sum_{n=(M+1)/2}^{(N-1)/2} \frac{P_{-n}}{m}\frac{\partial}{\partial Q_{-n}}+\frac{P_n}{m}\frac{\partial}{\partial Q_n} \no\\
 & - \frac{4}{\hbar}  U(\bQ)
 \sin\!\left({{\hat X}\over 2}\right)
 \cos\!\left({\hat Y}+{{\hat X}\over 2}\right)
\end{align}
with
\begin{align}
{\hat X}={\hbar\over 2}\sum_{n=(M+1)/2}^{(N-1)/2} \frac{\ola \partial}{\partial Q_{-n}}\frac{\ora \partial}{\partial P_{-n}}+
 \frac{\ola \partial}{\partial Q_n}\frac{\ora \partial}{\partial P_n}
\end{align}
and 
\begin{align}
{\hat Y}={\hbar\over 2}\sum_{n=-(M-1)/2}^{(M-1)/2} \frac{\ola \partial}{\partial Q_n}\frac{\ora \partial}{\partial P_n}
\end{align}

\section{Derivation of Matsubara time-correlation function}

To obtain the expression for $C_{ AB}^{[M]}(t)$ in  \eqn{mattsc}, we note that $B({\bf  P},{\bf Q},t)$ is independent of the non-Matsubara ${\bf P}$ modes (since, by construction,  these modes are not involved in the Matsubara dynamics) which can therefore be integrated out, giving a product of Dirac delta-functions in the non-Matsubara ${\bf \wt D}$ modes.\cite{pnote}  As a result, the Wigner transform $\left[ e^{-\beta{\hat H}} {\hat A}\right]_{\overline N}$ in \eqn{blobby} reduces to
\begin{align}
& \left[ e^{-\beta{\hat H}} {\hat A}\right]_{\overline N} ({\bf P}_M,{\bf Q}) \no\\
& = (2\pi\hbar)^{N-M}A({\bf Q})\int d{\bf D}_M \prod_{n=-(M-1)/2}^{(M-1)/2}e^{i{ P}_n D_n/\hbar} \no\\
& \ \ \   \times \prod_{l=1}^N \bra{\eta^-_{l-1}({\bf Q},{\bf D}_M)} \ebN \ket{\eta^+_l({\bf Q},{\bf D}_M) }
\end{align}
where ${\bf P}_M$ and ${\bf D}_M$ include only the Matsubara modes (and ${\bf Q}$  includes all $N$ modes), and
\begin{align}
\eta^\pm_l({\bf Q},{\bf D}_M)=\sum_{n=-(N-1)/2}^{(N-1)/2}T_{ln}Q_n \pm \sum_{n=-(M-1)/2}^{(M-1)/2}T_{ln}D_n/2
\end{align}
(where the  dependence of $\eta^\pm_l$ on $({\bf Q},{\bf D}_M)$ will be suppressed in what follows).
Expressing the bra-ket in ring-polymer form, and using trigonometric identities, we obtain
\begin{align}
& \left[ e^{-\beta{\hat H}} {\hat A}\right]_{\overline N} ({\bf P}_M,{\bf Q}) \no\\
& = (2\pi\hbar)^{N-M}\left({m\over 2\pi\beta_N\hbar^2}\right)^{N/2}A({\bf Q}) \int d{\bf D}_M\no\\
& \ \ \ \times e^{-\beta_Nm  f_M({\bf Q},{\bf D}_M)/2} \prod_{n=-(M-1)/2}^{(M-1)/2}e^{i{ P}_n D_n/\hbar} \no\\
& \ \ \ \times   \exp\left\{-{\beta_N\over 2}\left[ \sum_{l=1}^N    V(\eta^-_{l})   +
V(\eta^+_{l}) \right]  \right\}                          
  \label{b3}
\end{align}
where 
\begin{align}
& f_M({\bf Q},{\bf D}_M) \no\\
& = {4\over (\beta_N\hbar)^2}\sum_{n=-(M-1)/2}^{(M-1)/2} \left(Q_n\sin{n\pi\over N}+{D_{-n}\over 2}\cos{n\pi\over N}
\right)^2\no\\
& \ \ \ + \sum_{n=(M+1)/2}^{(N-1)/2}(Q_n^2+Q_{-n}^2)\omega_n^2\label{b4}
\end{align}

On taking the limit \ninc and converting ${\bf D}_M$ to ${\bf \wt D}$,  we find that the Gaussians involving ${\bf D}_M$ in \eqn{b3} have the form
\begin{align}
\exp\left(-m{\wt D}_n^2 N^2/2\beta\hbar^2  \right)
\end{align}
i.e.\ each Gaussian in ${\bf \wt D}$ becomes a Dirac delta-function in the limit $N\to\infty$. This allows us to replace the third line in
\eqn{b3} by
\begin{align}
 \exp\left[-\beta_N \sum_{l=1}^N    V\left(\sum_{n=-(N-1)/2}^{(N-1)/2}T_{ln}Q_n\right)  
  \right]     
\end{align}
and to integrate out the ${\bf \wt D}$, giving
\begin{align}
& \left[ e^{-\beta{\hat H}} {\hat A}\right]_{\overline N} ({\bf P}_M,{\bf Q}) \no\\
& = \left({2\pi m\over \beta_N}\right)^{(N-M)/2}A({\bf Q}) \no\\
& \ \ \ \times e^{-\beta_N {\bf  P}_M^2/2m}
 \prod_{n=-(M-1)/2}^{(M-1)/2}e^{2i{ P}_n  {Q}_{-n}\tan(n\pi/N)/\hbar} \no\\
& \ \ \ \times \exp\left[-{\beta_N\over 2} \sum_{l=1}^N    V\left(\sum_{n=-(N-1)/2}^{(N-1)/2}T_{ln}Q_n\right)   \right]    \no\\
& \ \ \ \times \exp\left[-{\beta_Nm\over 2}\sum_{n=(M+1)/2}^{(N-1)/2}(Q_n^2+Q_{-n}^2)\omega_n^2   \right]
\end{align}
We then substitute this expression into the integral of \eqn{blobby} (with $\int d{\bf P}$ replaced by $\int d{\bf P}_M$), and take the limit
 $M\to\infty$ (subject to $M\ll N$), which allows us to integrate out the non-Matsubara modes in ${\bf Q}$. Use of the formula\cite{grad}
 \begin{align}
\prod_{n=1}^{N-1}\sin\left(n\pi/ N\right)={N/2^{N-1}}
 \end{align}
 then gives \eqnt{mattsc}{thet}.


\begin{thebibliography}{10}
\bibitem{def} This definition of the time-correlation function (excluding the $1/Z$) will be used throughout.
\bibitem{summar} This paragraph is a heuristic summary of what is derived properly in refs.\ \onlinecite{miller_molphys} and \onlinecite{shi_geva}.
\bibitem{even} All the derivations reported here can also be done for even $N$, at the cost of doubling the amount of algebra in order to deal with the awkward $N/2$-th normal mode.
\bibitem{arti} Clearly this distinction is artificial, since, for any $M\ll N$, there will be `non-Matsubara' modes for which $|n|>(M-1)/2$ and $|n|\ll N$, and which therefore also satisfy \eqn{matty}. However, all we need to know is that all of the $M$ Matsubara modes become increasingly smooth in the limit $N\to\infty$, and that the majority of the $N\!-\!M$ non-Matsubara modes do not.
\bibitem{der2} We could thus have derived Matsubara dynamics by starting with the LSC-IVR Liouvillian ${\cal L}_{N}$ of \eqn{norml}, then discarding the non-Matsubara derivative terms; but this would have hidden the important property that Matsubara dynamics is inherently classical.
\bibitem{sc-ivr} The persistence of the oscillations in the Matsubara result for the quartic oscillator (Fig.~6a) suggests that  a Matsubara version of (non-linearized) semiclassical-IVR (i.e.\ the next approximation up in the hierarchy, in which individual forward-backward paths are treated classically and assigned phases\cite{miller_rev}) may give very close agreement with the exact quantum result for significantly longer times.
\bibitem{high} These two approximations are identical only in the limits in which LSC-IVR and Matsubara dynamics agree and are both exact (i.e.\ the short-time, harmonic and high-temperature limits).
\bibitem{pnote} Note that if ${\hat B}$ is a function of the momentum operator, then $B({\bf  P},{\bf Q},t)$ does depend on the non-Matsubara ${\bf P}$ coordinates, but that this dependence has a known, analytic, form, such that these coordinates can be integrated out, converting the original dependence on the non-Matsubara ${\bf P}$ coordinates into a dependence on the non-Matsubara ${\bf Q}$ coordinates.
\end{thebibliography}
\end{document}